\documentclass[aps,twocolumn,pra,amssymb,showpacs]{revtex4}

\usepackage[dvips]{graphicx}
\def\mathvecfont#1{\textbf{\em #1}}
\newcommand{\myvec}[1]{\mathvecfont{#1}}
\newcommand{\ci}{\mathrm{i}}

\usepackage[usenames,dvipsnames]{xcolor}
\usepackage[colorlinks=true, linkcolor=RoyalBlue, citecolor=Orange]{hyperref}

\newcommand{\subref}[2]{\ref{#1}{{\color{RoyalBlue}\,(#2)}}}
\newcommand{\sub}[1]{{#1}}
\newlength{\figwidth}
\newcommand{\eref}[1]{(\ref{#1})}
\begin{document}

\title{%
Probing the Band Topology of Mercury Telluride\\ through
Weak Localization and Antilocalization\\[0.3cm]
}

\author{Viktor Krueckl}
\address{Institut f\"ur Theoretische Physik, Universit\"at Regensburg, D-93040 Regensburg, Germany}

\author{Klaus Richter}
\address{Institut f\"ur Theoretische Physik, Universit\"at Regensburg, D-93040 Regensburg, Germany}

\begin{abstract}
We analyze the effect of weak localization (WL) and weak antilocalization (WAL) in the
electronic transport through HgTe/CdTe quantum wells.
We show that for increasing Fermi energy
the magnetoconductance  of a diffusive system with inverted band ordering features a
transition from WL to WAL and back, if spin-orbit interactions 
from bulk and structure inversion asymmetry can be neglected. 
This, and an additional splitting in the magnetoconductance profile, is a signature of the
Berry phase arising for inverted band ordering and not present in heterostructures with
conventional ordering.
In presence of spin-orbit interaction both band topologies 
exhibit WAL, which is distinctly energy dependent solely for quantum wells with inverted band ordering.
This can be explained by an energy-dependent decomposition of the Hamiltonian
into two blocks.
\end{abstract}

%Uncomment for PACS numbers title message
%\submitto{\SST}

\maketitle

\section{Introduction}

The first theoretical proposal for a two-dimensional topological insulator
was based on graphene with intrinsic spin-orbit interaction (SOI)~\cite{Kane2005b,Kane2005}.
Although the spin-orbit coupling of graphene is too small
to render an experimental evidence~\cite{Huertas-Hernando2006, Gmitra2009},
this initiated several other suggestions for two-dimensional materials and heterostructures 
showing topological insulator features~\cite{Bernevig2006b, Murakami2006, Liu2008}.
Subsequently, the criteria for topological insulators were extended to three dimensions~\cite{Fu2007b}
and were experimentally verified in other suitable materials like BiSe~\cite{Fu2007} 
or Bi$_2$Te$_3$~\cite{Qu2010, Xiu2011}.
In the meantime, the quantum spin Hall effect, a prominent transport feature of
two-dimensional topological insulators, has been observed in
HgTe/CdTe quantum wells~\cite{Koenig2007, Roth2009} 
as well as for InAs/GaSb heterostructures~\cite{Knez2011}.
In both experiments the transmission through a mesoscopic Hall bar is quantized since the
bulk of the system is insulating and the current is only carried by edge states,
which are protected from backscattering due to time-reversal symmetry.

Whilst many theoretical investigations are focused on these edge
states of two-dimensional topological insulators~\cite{Zhou2008, Zhang2011, Krueckl2011b, Dolcini2011, Budich2012},
signatures of the special band topology are also traceable in other observables
even away from the bulk band gap.
To this end we consider a well studied phenomenon in phase coherent transport through disordered
quantum systems, namely weak localization (WL)~\cite{Altshuler1980} for systems without SOI 
and weak antilocalization (WAL)~\cite{Hikami1980} in presence of SOI. 
The effect stems from the self interference of the charge carriers leading to
a quantum correction to the classical transmission for time reversal symmetric systems.
Breaking of this symmetry can be easily achieved by applying a perpendicular magnetic field.
In a semiclassical picture, the effect is understood in terms of interference between
waves traveling in opposite directions along backscattered paths and
averaging over all such trajectory pairs. 
Besides the relative phase shift arising from the enclosed flux of an external
perpendicular magnetic field, intrinsic Berry phases affect the 
interference and thereby the WL behavior. 
As a consequence, the signatures of WL in transport through systems
with strong Berry phases, as for example HgTe heterostructures,
can differ significantly from those of conventional electron gases.

To our knowledge, there are only a few theoretical studies of WL in systems
with inverted band ordering~\cite{Tkachov2011, Lu2011PRL}.
Diagrammatic studies for the two-dimensional case show a transition between
WL and WAL upon varying the chemical potential~\cite{Tkachov2011}, similar
to the WL physics in topological insulator thin films~\cite{Lu2011PRB, Garate2012},
which is supported by experimental investigations~\cite{He2011, Liu2012b}.
However, major SOI effects from bulk and inversion asymmetry are not included, which alter the
WL signal, as we will show in this work.
A recent experiment with HgTe heterostructures revealed WAL in diffusive transport~\cite{Olshanetsky2010}
and detailed investigations attested an energy dependence of the WAL peak~\cite{Minkov2012}.
Since no theories for WL in heterostructures with inverted band ordering including SOI 
are at hand, only conventional theories
for A$_3$B$_5$ semiconductors~\cite{Iordanskii1994,Knap1996} have yet been applied
to analyze these results.

In order to explore how WL effects are altered by the
inverted band ordering of topological insulators, we perform numerical transport calculations.
We confirm the transition between WL and WAL reported in Ref.~\cite{Tkachov2011},
if SOI can be neglected.
We explain the effect in terms of the Berry phase of the bands involved.
Moreover, we additionally find a splitting of the WL magnetoconductance profiles
due to the two spin species that can also be traced back to the Berry phase and is
not accounted for in previous diagrammatic studies.
Additionally, we show how the WL phenomenon is altered by SOI, 
and how bulk and structure inversion asymmetry 
lead to significantly different WAL features that can strongly depend on the band ordering.

This paper is structured as follows:
In Section~\ref{sec:model}
we introduce the model used to describe multi-band quantum transport in diffusive HgTe heterostructures.
In Section~\ref{sec:noso} we focus on effects of the Berry phase and the energy dependence of WL and WAL
without SOI. 
In Section~\ref{sec:withso} we include SOI 
and show why a variation in WAL upon change in energy serves as an indicator for 
inverted band ordering.
Finally, in Section~\ref{sec:conclude} we conclude with a brief summary.

\section{Model}
\label{sec:model}

We describe the electronic properties of the underlying HgTe heterostructure by the
Hamiltonian~\cite{Bernevig2006b, Rothe2010}
%
%
%%%%%%%%%%%%%%%
\begin{equation}
H = \left ( \begin{array}{c c c c}
C_k + M_k & A k_+ & - \ci R k_- & -\Delta \\
A k_- & C_k  - M_k & \Delta & 0 \\
\ci R k_+ & \Delta&  C_k  + M_k  & -A k_-\\
-\Delta& 0 & -A k_+& C_k - M_k\\
\end{array}
\right )
\label{Hhgte}
\end{equation}
%%%%%%%%%%%%%%%
%
%
where $k_\pm = k_x \pm \ci k_y$, $\myvec k^2 = k_x^2 + k_y^2$, $C_k = D \myvec k^2$ and $M_k = M - B \myvec k^2$.
The material parameters are chosen to be 
$A=354.5\,\mathrm{meV\,nm}$,
$B=-686\,\mathrm{meV\,nm}$,
%$C=0$,
$D=-512\,\mathrm{meV\,nm^2}$
and $M=\pm10\,\mathrm{meV}$.
Without SOI ($R = \Delta = 0$) this Hamiltonian breaks up into two independent 
${2}{\times}{2}$ blocks, each consisting of an $s$-like electron and a $p$-like hole band.
The topology of the band structure depends on the ordering of the electron and
hole states, given by the gap $M$ which is positive for conventional ordering ($M>0$) 
and negative for inverted ordering ($M<0$) .
Additionally, in Section~\ref{sec:withso} we take into account the SOI of strength $\Delta$ and $R$ due to bulk
inversion asymmetry (BIA) as well as
structure inversion asymmetry (SIA)~\cite{Rothe2010}.
While $\Delta$ is fixed (we use $\Delta = 1.6\,\mathrm{meV}$~\cite{Koenig2008}),
the strength of the SOI due to  SIA depends on the quantum well structure, and can be 
tuned to very small values by
growing symmetric wells.

\begin{figure}[tb]
\centering
\includegraphics[width=\figwidth]{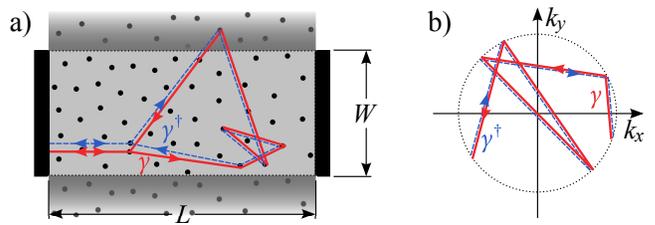}
\caption{
\sub{a})~Sketch of the scattering region with periodic boundary conditions in vertical
direction between two non-periodic leads.
A typical backscattered path and its time reversed counter path are shown, 
contributing to WL and WAL.
\sub{b})~Corresponding momentum-space trajectory for the two paths of (a).
}
\label{figsketch}
\end{figure}

We study the signatures of WL in magnetotransport 
through diffusive conductors in the presence of a perpendicular magnetic field $B$.
We consider coherent two-terminal transport through disordered strip geometries
with a Gaussian correlated disorder,
%
%
%%%%%%%%%%%%%
\begin{equation}
 U(\myvec r) = \sum_i U_i \exp\left({- \frac{(\myvec r - \myvec R_i)^2}{2 \sigma^2} }\right) \, ,
 \label{impuritypotential}
\end{equation}
%%%%%%%%%%%%%
%
%
with a correlation length $\sigma$.
Here, a box distribution, $ -U_0 \leq U_i \leq U_0$, is chosen
for the strength $U_i$ of the impurity $i$ located at $\myvec R_i$.
In order to eliminate the influence of the edge states we employ periodic
boundary conditions in the scattering region, linking the upper and lower 
edges along transport direction as sketched in Fig~\subref{figsketch}{a}.
%
%
%%%%%%%%%%%%%
%%%%%%%%%%%%%
%
%
We discretize the Hamiltonian (\ref{Hhgte}) on a square grid with a lattice
spacing of $5\,\mathrm{nm}$.
The conductance 
%
%
%%%%%%%%%%%%%
\begin{equation}
G=\frac{e^2}{h} T
 =\frac{e^2}{h}\sum_{n,m}\sum_{\sigma, \sigma'}
\vert t_{m, \sigma'; n, \sigma} \vert^2 
\end{equation}
%%%%%%%%%%%%%
%
%
is calculated in linear response within the Landauer formalism~\cite{Landauer1970},
whereby the transmission amplitudes $t_{m, \sigma'; n, \sigma}$
are given by the Fisher-Lee relations~\cite{Fisher1981}.
The indices $m$ and $n$ stand for the different modes in the leads, which are additionally 
classified through the index $\sigma \in \{ \mathrm{U}, \mathrm{L} \}$
denoting the upper left $(\mathrm{U})$ and the lower right $( \mathrm{L})$ block of
the Hamiltonian~\eref{Hhgte} if no SOI is present ($\Delta=R=0$).

\section{Berry phase effects in quantum transport without spin-orbit interaction}
\label{sec:noso}

In the following, we assume a negligibly small BIA and SIA spin-orbit
interaction leading to a Hamiltonian \eref{Hhgte} with two uncoupled blocks.
We will show that the Berry phase of each of those blocks leads to an
energy dependence of the WL signal different for the 
two band orderings.
Without losing generality we focus on the upper subblock
\begin{equation}
H_U = \left ( \begin{array}{c c}
M-(B+D) \myvec k^2 & A (k_x + \ci k_y)\\
A (k_x - \ci k_y) & -M+(B-D) \myvec k^2\\
\end{array}
\right ),
\label{Hup}
\end{equation}
since the results for the lower subblock can be obtained by applying the time reversal operator.
This Hamiltonian can be easily diagonalized, leading to the 
energy dispersion for the electron and hole branch,
\begin{equation}
E_{e/h}(\myvec k) =
- D \myvec k^2 \pm F(\myvec k), 
\end{equation}
with
\begin{equation}
F(\myvec k) =  \sqrt{A^2 \myvec k^2 + ( B \myvec k^2 - M )^2} .
\end{equation}
The corresponding eigenstates are given by
\begin{equation}
\psi_{e/h}(\myvec k) \propto \left ( \begin{array}{c}
M - B \myvec k^2 \pm F(\myvec k)\\
A (k_x - \ci k_y)
\end{array} \right).
\end{equation}
For vanishing SOI, the WL properties are governed by the phases accumulated
by one of these spinors.
In a semiclassical description quantum corrections to the conductance stem
from the interference of waves traveling along different impurity-scattered
paths.
Upon disorder average the contributions from pairs of uncorrelated paths vanish.
The remaining contributions leading to WL mainly originate from pairs of a path $\gamma$
with its time inverted path $\gamma^\dagger$ where the dynamical phases
cancel out, as depicted in Fig.~\subref{figsketch}{a}.
As a result, the WL signal is governed by additional phases,
like the phase due to the flux of an external magnetic field or a Berry phase.
The latter is associated with the Berry curvature given by~\cite{Berry1984, Chang1996}
\begin{equation}
\mathcal{A}_\sigma(\myvec{k}) =
    -\ci\langle \psi_\sigma(\myvec k) \vert \nabla_k \psi_\sigma(\myvec k) \rangle \, ,
\end{equation}
in terms of the bulk eigenstates $\psi_\sigma(\myvec k)$.
The corresponding phase is obtained by integrating the vector potential $\mathcal{A}_\sigma$
along a backscattered path corresponding to a closed loop in momentum space with a fixed
momentum $k = \vert \myvec k \vert $,
as sketched in Fig.~\subref{figsketch}{b}:
\begin{equation}
\Gamma_\sigma = 
\oint_{k=\mathrm{const}} \hspace{-0.7cm}
\mathcal{A}_\sigma(\myvec{k}) \cdot \mathrm{d}\myvec{k}
= 2 \pi \mathcal{A}_{\sigma}(\myvec{k}) \cdot (-k_y, k_x) \, .
\label{berryphaseangle}
\end{equation}
Because of the circular symmetry of $\mathcal{A}_\sigma(\myvec{k})$ the phase
$\Gamma_\sigma$ can be evaluated by the scalar product 
$\mathcal{A}_{\sigma}(\myvec{k}) \cdot (-k_y, k_x)$ at a
single point in momentum space.
This geometric phase $\Gamma_\sigma$ enters the semiclassical Greens function.
%
%
%%%%%%%%%%%%%
\begin{figure}[tb]
\centering
\includegraphics[width=\figwidth]{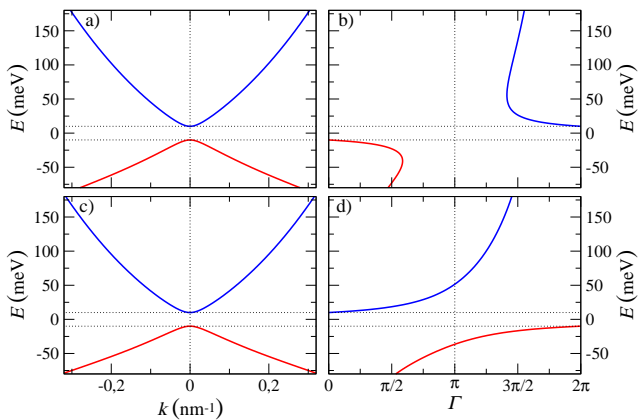}
\caption{
Bulk band structure (\sub{a}),(\sub{c}) and corresponding Berry phase Eq.~\eref{berryphaseangle}.
(\sub{b}),(\sub{d}) of the Hamiltonian \eref{Hhgte}.
Top panels show the result for conventional band ordering ($M = 10\,\mathrm{meV}$),
bottom panels the result for inverted band ordering ($M =-10\,\mathrm{meV}$).
}
\label{figberry}
\end{figure}
%%%%%%%%%%%%%
%
%
As depicted in Fig.~\subref{figsketch}{b}, a backscattered path and its time-inverted partner accumulate
opposite reflection angles in momentum space.
In view of Eq.~\eref{berryphaseangle}, this opposite angle leads to opposite Berry phases 
and thereby to a dephasing  in the two-path interference.
This results in a reduction of WL~\cite{krueckl2011a}, 
right up to a complete reversal of the WL correction
to full WAL~\cite{Suzuura2002, McCann2006a}.
For the Hamiltonian~\eref{Hhgte} the geometric phase 
 $\Gamma_\sigma$ has remarkable properties depending on the two different
band topologies.
In the case of HgTe the strength of the geometric phase of the electron and the
hole branch are given by
\begin{equation}
\Gamma_{e/h} = \pi
\left (
1 \pm 
\frac{M-B \myvec{k}^2}{F(\myvec{k})}
\right ).
\end{equation}
Although the band structure of the conventional ($M>0$) and inverted ($M<0$) ordering is very
similar [compare Fig.~\subref{figberry}{a),(c}], the Berry phases of the different systems are not.
For the inverted band ordering, the Berry phase spans the whole range of possible 
phases from $0$ to $2\pi$, as shown in Fig.~\subref{figberry}{d}.
As a consequence, a particular energy exists where the accumulated phase
$\Gamma_\sigma=\pi$, as in a ``neutrino billiard"~\cite{Berry1987}.
However, if the bands are ordered conventionally, this is not the case.
Although the phase differs  significantly from $0$ or $2\pi$,
the region around $\pi$ is excluded as shown in Fig.~\subref{figberry}{b}.
As a consequence, we expect distinctly different WL behavior for both systems
if the whole energy range is considered.

In the following, we study the WL correction  in transport 
through a disordered HgTe heterostructure numerically
by calculating the average change of the quantum transmission
%
%
%%%%%%%%%
\begin{equation}
\delta T(B) = \big\langle T(B) - T(0) \big\rangle
\label{deltaT}
\end{equation}
%%%%%%%%%
%
%
in presence of a perpendicular magnetic field $B$.
We tune the Berry phase by changing the Fermi energy of the system.
The averages are taken over a set of $1000$ different impurity potentials \eref{impuritypotential}
distributing $20000$ impurities (equals a coverage of $10\%$ of the grid points) within a scattering region  of 
$1000\,\mathrm{nm} \times 5000\,\mathrm{nm}$ with a correlation length $\sigma = 15\,\mathrm{nm}$.
The disorder strength $U_0$ is tuned to get a constant mean free path of $1200\,\mathrm{nm}$
for all energies, leading to comparable shapes of the localization correction for a large range of Fermi energies.

%
%%%%%%%%%%%%%
\begin{figure}[tb]
\centering
\includegraphics[width=\figwidth]{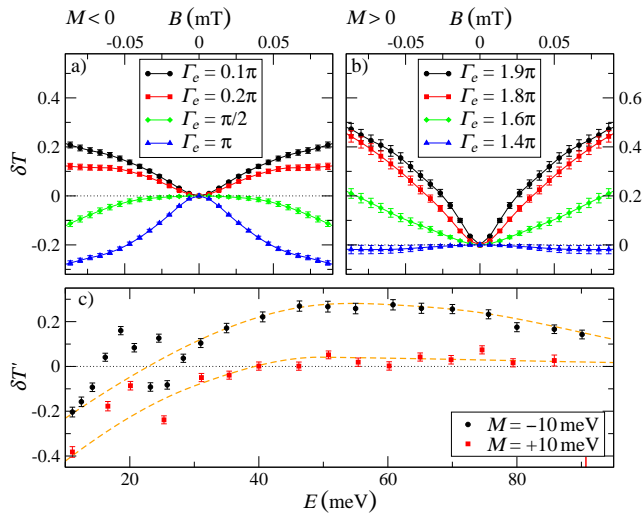}
\caption{
Weak localization correction in a disordered  HgTe nanoribbon ($L=5000\,\mathrm{nm}$, $W=1000\,\mathrm{nm}$).
Upper panels: Magnetic field dependence for (a) inverted band ordering and
(b) conventional ordering.
Different Fermi energies ($E_F\,{=}\,\{11.1\,\mathrm{meV}, 12.5\,\mathrm{meV}, 18.5\,\mathrm{meV}, 52\,\mathrm{meV}\}$
 from top to bottom)
lead to a Berry phase as given in panel (a) and (b).
Impurity potential strength $U_0$ is varied to fix a mean free path of $1200\,\mathrm{nm}$.
\sub{c}) Energy dependence of the WL correction $\delta T'$, Eq.~\eref{deltaTp}, for 
inverted and conventional ordering extracted for a magnetic field 
$B=0.1\,\mathrm{mT}$.
Dashed curves are guides to the eye.
}
\label{figlocnoso}
\end{figure}
%%%%%%%%%%%%%
%
%
The results are summarized in Fig.~\ref{figlocnoso}.
For energies close to the band gap, the Berry phase is very small in both cases.
As a result, the average transmission is similar to that of an electron gas, leading to 
WL, visible as a negative correction to the magnetotransmission and
shown as black line in Fig.~\subref{figlocnoso}{a} for the case of inverted band ordering.
By increasing the Fermi energy also the Berry phase raises, leading to a reduced
WL correction.
For values of $\Gamma_\sigma = \pi/2$, the minimum in the average transmission
at $B=0$ is expected to vanish, which is also reflected in the numerical data presented as 
green line in Fig.~\subref{figlocnoso}{a}.
If the energy is tuned to 
%
%
%%%%%%
\begin{equation}
E^{(\pi)}_e = -\frac{D M}{B} + \sqrt{\frac{A^2 M}{B}} ,
\label{topoEpi}
\end{equation}
%%%%%%
%
%
such that the momentum $\myvec k$ fulfills $B \myvec{k}^2 = M$, the regime 
with a Berry phase close to $\pi$ is entered.
In this configuration, the system is expected to feature WAL, since a path and
the time inverted counter path accumulate a phase difference of $\pi$
and therefore interfere destructively, leading to an enhanced transmission at $B=0$.
This is indeed visible in the numerical results (Fig.~\subref{figlocnoso}{a} as blue line)
showing a pronounced WAL peak.

The physics changes, if a heterostructure with conventional ordering
of the quantum well states is considered.
In Fig.~\subref{figlocnoso}{b} we show the average magnetoconductance for the same configuration
as in Fig.~\subref{figlocnoso}{a}, however, we assume a positive bandgap of $M=10\,\mathrm{meV}$.
For Fermi energies close to the bandgap, the Berry phase is small, as in the case with
inverted band ordering, leading to a conventional WL feature.
Unexpectedly, the strength of the WL correction of the conventional
regime is almost twice as strong as the result for the inverted regime
[compare black lines in Figs.~\subref{figlocnoso}{a,b}].
With increasing Fermi energy, the strength of the Berry phase increases, but does
not reach $\pi$.
Instead, the maximal phase at $B \myvec{k}^2 = M$ is rather close to $\pi/2$, leading to a
strong reduction of any localization correction [see blue line of Fig.~\subref{figlocnoso}{b}].

For a more closer analysis of the energy dependence we extract the strength of the WL
correction,
%
%
%%%%%%%%%
\begin{equation}
\delta T' = \big\langle T(B=0) - T(B=0.1\,\mathrm{mT})  \big\rangle ,
\label{deltaTp}
\end{equation}
%%%%%%%%%
%
%
for various Fermi energies.
The results are summarized in Fig.~\subref{figlocnoso}{c}.
For conventional ordering, we get a transition from WL close to the band gap 
to almost no localization for higher energies.
For inverted band ordering one finds a clear-cut transition, from WL to WAL and back to WL.
Note that for very low energies only few channels contribute to transport.
As a consequence the strength of the WL correction is reduced
due to the finite number of open channels~\cite{Beenakker1997}, and non-universal 
features may appear.
These apparently erratic values vanish when the width of the scattering region  is increased.

Additional to the expected crossover from WL to WAL, the Berry phase leads furthermore
to opposite shifts in $B$ of the magnetotransmission profiles associated with the upper and lower
blocks of the Hamiltonian~\eref{Hhgte}.
A pair of backscattered paths, contributing to WL and WAL, can be characterized
in terms of the enclosed area $A$ and the accumulated angle $\alpha$,
acquired during the series of random scattering processes at impurities along the
diffusive path.
Usually, only the enclosed areas $A$ are relevant, and their typical value $A_0$ 
sets the magnetic field scale in the magnetoconductance profile;
{\em i.e.} its width is of order $B A_0 \propto \Phi_0$, where $\Phi_0$ is 
the magnetic flux quantum.
However, as has been recently shown for ballistic and diffusive 
two-dimensional hole gases (based on the ${4}{\times}{4}$ Luttinger Hamiltonian)~\cite{krueckl2011a},
an underlying Berry phase gives rise to a characteristic shift of
the WL peak.
This shift depends on the associated Berry phase $\Gamma$ and the
typical accumulated angle $\alpha_0$.
Moreover, for diffusive and chaotic conductors 
there is a finite classical correlation $\rho$ between the random variables $A$ and $\alpha$.
These different quantities determine an effective magnetic ``Berry field" $\tilde B$
by which the WL magneto profile is shifted.
For a chaotic quantum dot, this shift corresponds to an associated flux~\cite{krueckl2011a}
%
%
%
%%%
\begin{equation}
  \tilde B\,A_0 \propto \left ( \Gamma \, \rho \, \frac{\alpha_0 }{2 \pi } \right ) \Phi_0 ,
  \label{effectiveBerryfield}
\end{equation}
%%%
%
%
which depends linearly on the Berry phase $\Gamma$, the typical enclosed angle $\alpha_0$
and the classical correlation $\rho$.
This behavior has also been found for ballistic cavities based on HgTe~\footnote{%
See supplementary material of Ref.~\cite{krueckl2011a}.}.
%
%
%%%%%%%%%%%%%
\begin{figure}[tb]
\centering
\includegraphics[width=\figwidth]{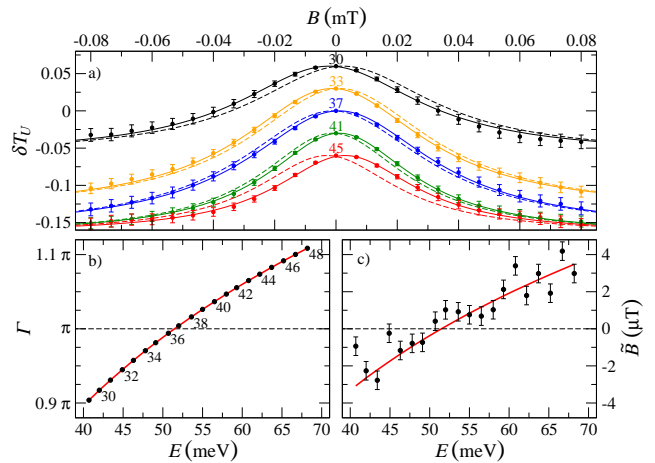}
\caption{
$B$-field shift of the WAL maximum due to correlations between
enclosed area and angle of the contributing trajectories.
\sub{a}) Magnetic field dependence of transmission quantum correction
of a diffusive periodic nanoribbon ($W=1000\,\mathrm{nm}$, $L=5000\,\mathrm{nm}$)
for different number of open channels ($30$ to $45$) close to the
Fermi energy $E_F = 52\,\mathrm{meV}$
(displayed with vertical offset).
Symbols with error bars: Results $\delta T_U$ for upper block, see Eq.~\eref{deltaT};
solid lines: fit to numerical data; dashed lines: corresponding
curve for $\delta T_L$ of lower block.
\sub{b}) Energy dependence of Berry phase around  $52\,\mathrm{meV}$ (see Fig.~\subref{figberry}{a}).
\sub{c}) Energy dependence of shift $\tilde B$ of the magnetotransmission
maximum  extracted from (\sub{a}).
}
\label{figshift}
\end{figure}
%%%%%%%%%%%%%
%
%
For the disordered HgTe quantum well, we expect a corresponding behaviour, not only for the WL,
but also for the WAL peaks.

Fig.~\subref{figshift}{a} shows the numerically obtained quantum correction to the
magnetoconductance $\delta T_\mathrm{U}$
(bullets) based on the upper block $\mathrm{U}$ of the Hamiltonian~\eref{Hhgte}.
The five different curves correspond to different Fermi energies, close to
$E_F=52\,\mathrm{nm}$, labeled by the number of open transverse modes (without spin)
varying from 30 to 45.
Fits to the numerical data are shown as solid lines.
Correspondingly, the dashed lines show the further contribution from the lower block $\mathrm{L}$.
The curves exhibit a small but distinct energy-dependent shift in $B$, respectively, a splitting
of the magnetoconductance of different blocks.
This feature can be explained in terms of the Berry field introduced above.
To this end, the Berry phase $\Gamma$ corresponding to the Fermi energy $E$,
respectively,  a number of open modes is shown in Fig.~\subref{figshift}{b}.
Since $\Gamma$ is close to $\pi$ in the energy range shown, all magnetoconductance
curves show WAL.
Most notably, the sign change in $\Gamma\,{-}\,\pi$ between energies corresponding
to 36 and 37 channels in Fig.~\subref{figshift}{b} is reflected in the 
direction of the energy dependent shift of the WAL curves in Fig.~\subref{figshift}{a},
with a nearly vanishing shift close to the trace with $n=37$.
Hence, Fig.~\subref{figshift}{b} illustrates the transition from negative correction
to positive correction between 36 and 37 open channels.
The same transition is also visible in the effective Berry field $\tilde B$,
which we extracted for various magnetoconductance curves 
by the same fits as shown in Fig.~\subref{figshift}{a}.
The effective Berry field $\tilde B$ is depicted in Fig.~\subref{figshift}{c}.
In view of Fig.~\subref{figshift}{b} its energy dependence 
indicates a linear dependence on the Berry phase as
it is the case in chaotic quantum dots,
see Eq.~\eref{effectiveBerryfield}~\cite{krueckl2011a}.
Due to the relatively low correlation $\rho$ between $\alpha$ and $A$
for a diffusive process, we expect the strength of the shift to be
only a few $\mu T$ in the present case.
However, such a shift leads to a significant change of the magnetoresistance line shape.
To the best of our knowledge, it is not captured by any existing diagrammatic approach or 
description by random matrix theory.

\section{Role of spin-orbit interaction}
\label{sec:withso}

In addition to the spin-orbit coupling  between the $s$- and $p$-bands within the
${2}{\times}{2}$ blocks, there are further spin-orbit interactions present in HgTe heterostructures.
Those can be divided due to their physical origin into terms arising from bulk inversion
asymmetry (BIA) and  structure inversion asymmetry (SIA).
BIA is given by the crystal structure itself, and thus can only be modified by 
changing the material.
SIA depends on internal and external electric fields, and consequently 
changes its size depending on
the symmetries of the grown HgCdTe layers or external gating.
For symmetric HgTe quantum wells the strength of SIA is negligibly small. 

%
%
%%%%%%%%%%%%%
\begin{figure}[tb]
\centering
\includegraphics[width=\figwidth]{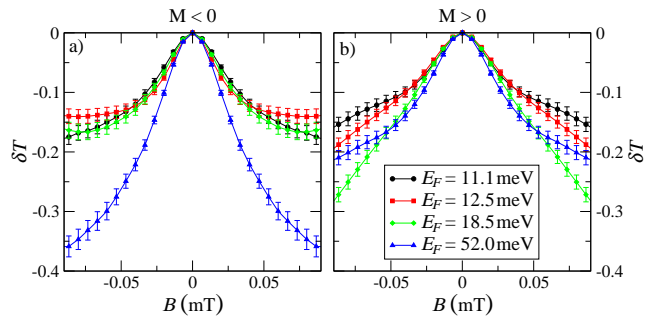}
\caption{
Strength of the WL correction $\delta T$ in presence of 
spin-orbit interaction due to structure inversion asymmetry
($\Delta=1.6\,\mathrm{meV}$) for 
(\sub{a}) inverted band order ($M=-10\,\mathrm{meV}$) and
(\sub{b}) conventional band order ($M=+10\,\mathrm{meV}$).
}
\label{figSIAonly}
\end{figure}
%%%%%%%%%%%%%
%
%
In the following, we first focus on the WAL profile in a
symmetric heterostructure with a naturally sized BIA and without SIA.
Our results for different Fermi energies are summarized in Fig.~\subref{figSIAonly}{a}
for inverted band ordering and in Fig.~\subref{figSIAonly}{b} for 
conventional band ordering.
The energies are chosen to cover the full range of Berry phases as in Fig.~\ref{figlocnoso}.
In comparison to systems without additional SOI the average
magnetoconductance always features WAL.
This is in line with the common explanation that strong SOI leads to spin relaxation and thereby WAL.
However, there exists a significant difference between the energy dependence of the
WAL strength for the different band orderings.
For conventional ordering, the WAL correction is almost constant
and also the shape of the correction does not change significantly with Fermi energy, as
shown in Fig.~\subref{figSIAonly}{b}.
This is not the case for the inverted ordering.
Here, the correction is almost twice as strong if the Fermi energy is chosen to be
$E^{(\pi)}_e=52\,\mathrm{meV}$, Eq.~\eref{topoEpi},
the point with a Berry phase of $\pi$, as depicted in  Fig.~\subref{figSIAonly}{a}.

In the following, we give an explanation for this difference.
To this end, we apply the unitary transformation
%
%
%%%%%%%%%%%%%
\begin{equation}
\mathcal{T}
=\frac{1}{\sqrt{2}} \left (
\begin{array}{c c c c}
1 & 0 & 0 & 1 \\
0 & -1 & 1 & 0 \\
-1 & 0 & 0 & 1 \\
0 & 1 & 1 & 0
\end{array} \right )
\end{equation}
%%%%%%%%%%%%%
%
%
to the Hamiltonian (\ref{Hhgte}), leading to the transformed Hamiltonian
%
%
%%%%%%%%%%%%%
\begin{equation}
H
= \left (
\begin{array}{c c c c}
C_k - \Delta & -A k_+ & -M_k & -\ci \frac{1}{2} R k_- \\
-A k_- & C_k - \Delta & -\frac{1}{2}\ci  R k_+ & M_k \\
-M_k & \frac{1}{2} \ci R k_- & C_k + \Delta & -A k_+ \\
\frac{1}{2} \ci R k_+ & M_k & -A k_- & C_k + \Delta
\end{array} \right ) .
\end{equation}
%%%%%%%%%%%%%
%
%
%
%
%%%%%%%%%%%%%
\begin{figure}[t]
\centering
\includegraphics[width=\figwidth]{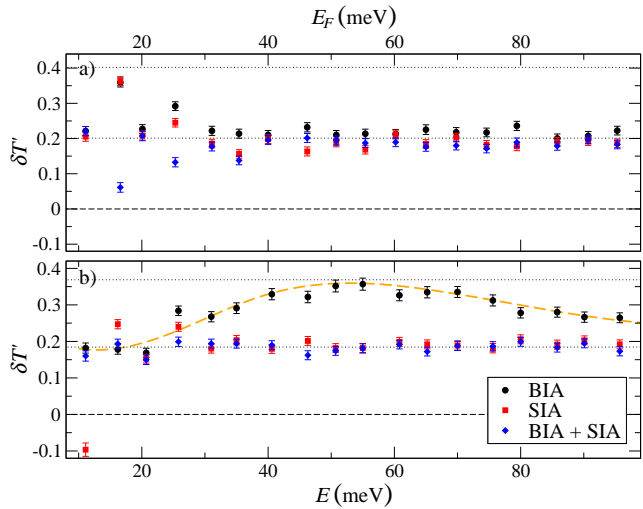}
\caption{
Strength of the WL correction $\delta T'$ for different spin-orbit
interactions  (BIA $\Delta=1.6\,\mathrm{meV}$, SIA $R=35\,\mathrm{eV\AA}$). 
Results are extracted from the transmission at $0.1\,\mathrm{mT}$.
\sub{a}) The localization strength for conventional band ordering ($M=10\,\mathrm{meV}$)
shows the same localization strength for all combinations of structure
and bulk inversion asymmetry.
\sub{b}) For inverted band ordering ($M=-10\,\mathrm{meV}$) and pure structure inversion asymmetry,
the strength of the WAL correction is doubled at $E^{(\pi)}_e = 52\,\mathrm{meV}$.
}
\label{figfourtypes}
\end{figure}
%%%%%%%%%%%%%
%
%
If no SOI %spin-orbit interaction
due to SIA is present ($R=0$), this Hamiltonian consists of two blocks which are only
coupled by the matrix element $M_k = M - B \myvec{k}^2$.
For an inverted band ordering there exists a momentum $\myvec k$, where 
$M_k \approx 0$ since $M$ and $B$ are both negative.
In HgTe superstructures with $M=-10\,\mathrm{meV}$ the 
Fermi energy is $52\,\mathrm{meV}$ corresponding to $E_e^{(\pi)}$.
At this energy the Hamiltonian decouples into two independent ${2}{\times}{2}$ blocks
that both show WAL.
Thus the entire WAL strength is twice as high
compared to other energies, as shown in Fig.~\subref{figSIAonly}{a}.

If additional spin-orbit terms from SIA are present, this unitary transformation into
two uncoupled blocks is not possible.
As a consequence, the WAL correction stays roughly constant throughout the whole
range of different Fermi energies. 
In Fig.~\ref{figfourtypes} we have summarized the behaviour of the WAL correction $\delta T'$,
Eq.~\eref{deltaTp}, for different combinations of BIA and SIA.
As expected, the WAL with conventional band ordering is
independent of the type of SOI, as shown in Fig.~\subref{figfourtypes}{a}.
However, this is not the case for a heterostructure with inverted band ordering.
If only BIA is present, WAL is approximately doubled at $52\,\mathrm{meV}$
and shows a smooth transition in between, see black symbols in Fig.~\subref{figfourtypes}{b}.
For finite SIA a block diagonalization is not possible,
and hence the WAL correction is constant, independent of whether additional BIA SOI
is present.
As in the calculations without SOI, the erratic fluctuations of the
WAL strength at low energies can be attributed to the correspondingly limited amount of open channels
in the numerical calculations.

\section{Conclusion}
\label{sec:conclude}
In this manuscript, we have analyzed the weak localization properties of HgTe heterostructures 
with different band topologies.
We revealed a transition between weak localization  and weak antilocalization for systems without 
spin-orbit interaction, which is only complete for systems with inverted band ordering
and can be related to the effect of the Berry phase.
This Berry phase, moreover, affects the magnetoconductance line shape:
Owning to correlations in the statistics of backscattered paths that depends on
the type of classical dynamics (diffusive, chaotic or regular in the 
ballistic case) the Berry phase implies a splitting of the magnetoconductance profile. 
Furthermore, we showed that the band ordering can be deduced from the energy dependence
of the weak antilocalization correction in presence of spin-orbit interaction
due to bulk inversion asymmetry:
If the heterostructure features an inverted band ordering, the correction strength is energy
dependent in contrast to a constant weak antilocalization strength for conventional band ordering.
This is explained by an energy-dependent separation into two uncoupled bocks.
Additional Rashba-type spin-orbit interaction from structure inversion asymmetry
again diminishes the energy dependence.

\section*{Acknowledgments}
This work is supported by Deutsche Forschungsgemeinschaft
(GRK 1570 and joined DFG-JST Forschergruppe Topological Electronics).
We thank I. Adagideli, E. Hankiewicz, G. Tkachov and M. Wimmer for useful conversations.

%\bibliographystyle{prsty}
%\bibliography{library.bib}

\begin{thebibliography}{10}

\bibitem{Kane2005b}
\textsc{C.~L. Kane} and \textsc{E.~J. Mele}, {\em {Quantum Spin Hall Effect in
  Graphene}}, {\href{http://dx.doi.org/10.1103/PhysRevLett.95.226801}{Phys.
  Rev. Lett.}} {\bf
  {\href{http://dx.doi.org/10.1103/PhysRevLett.95.226801}{95}}},
  {\href{http://dx.doi.org/10.1103/PhysRevLett.95.226801}{226801}}
  {\href{http://dx.doi.org/10.1103/PhysRevLett.95.226801}{(2005)}}.

\bibitem{Kane2005}
\textsc{C.~L. Kane} and \textsc{E.~J. Mele}, {\em {Z₂ Topological Order and
  the Quantum Spin Hall Effect}},
  {\href{http://dx.doi.org/10.1103/PhysRevLett.95.146802}{Phys. Rev. Lett.}}
  {\bf {\href{http://dx.doi.org/10.1103/PhysRevLett.95.146802}{95}}},
  {\href{http://dx.doi.org/10.1103/PhysRevLett.95.146802}{146802}}
  {\href{http://dx.doi.org/10.1103/PhysRevLett.95.146802}{(2005)}}.

\bibitem{Huertas-Hernando2006}
\textsc{D. Huertas-Hernando, F. Guinea} and \textsc{A. Brataas}, {\em
  {Spin-orbit coupling in curved graphene, fullerenes, nanotubes, and nanotube
  caps}}, {\href{http://dx.doi.org/10.1103/PhysRevB.74.155426}{Phys. Rev. B}}
  {\bf {\href{http://dx.doi.org/10.1103/PhysRevB.74.155426}{74}}},
  {\href{http://dx.doi.org/10.1103/PhysRevB.74.155426}{155426}}
  {\href{http://dx.doi.org/10.1103/PhysRevB.74.155426}{(2006)}}.

\bibitem{Gmitra2009}
\textsc{M. Gmitra, S. Konschuh, C. Ertler, C. Ambrosch-Draxl} and \textsc{J.
  Fabian}, {\em {Band-structure topologies of graphene: Spin-orbit coupling
  effects from first principles}},
  {\href{http://dx.doi.org/10.1103/PhysRevB.80.235431}{Phys. Rev. B}} {\bf
  {\href{http://dx.doi.org/10.1103/PhysRevB.80.235431}{80}}},
  {\href{http://dx.doi.org/10.1103/PhysRevB.80.235431}{235431}}
  {\href{http://dx.doi.org/10.1103/PhysRevB.80.235431}{(2009)}}.

\bibitem{Bernevig2006b}
\textsc{B.~A. Bernevig, T.~L. Hughes} and \textsc{S.-C. Zhang}, {\em {Quantum
  Spin Hall Effect and Topological Phase Transition in HgTe Quantum Wells}},
  {\href{http://dx.doi.org/10.1126/science.1133734}{Science}} {\bf
  {\href{http://dx.doi.org/10.1126/science.1133734}{314}}},
  {\href{http://dx.doi.org/10.1126/science.1133734}{1757}}
  {\href{http://dx.doi.org/10.1126/science.1133734}{(2006)}}.

\bibitem{Murakami2006}
\textsc{S. Murakami}, {\em {Quantum Spin Hall Effect and Enhanced Magnetic
  Response by Spin-Orbit Coupling}},
  {\href{http://dx.doi.org/10.1103/PhysRevLett.97.236805}{Phys. Rev. Lett.}}
  {\bf {\href{http://dx.doi.org/10.1103/PhysRevLett.97.236805}{97}}},
  {\href{http://dx.doi.org/10.1103/PhysRevLett.97.236805}{236805}}
  {\href{http://dx.doi.org/10.1103/PhysRevLett.97.236805}{(2006)}}.

\bibitem{Liu2008}
\textsc{C. Liu, T. Hughes, X.-L. Qi, K. Wang} and \textsc{S.-C. Zhang}, {\em
  {Quantum Spin Hall Effect in Inverted Type-II Semiconductors}},
  {\href{http://dx.doi.org/10.1103/PhysRevLett.100.236601}{Phys. Rev. Lett.}}
  {\bf {\href{http://dx.doi.org/10.1103/PhysRevLett.100.236601}{100}}},
  {\href{http://dx.doi.org/10.1103/PhysRevLett.100.236601}{236601}}
  {\href{http://dx.doi.org/10.1103/PhysRevLett.100.236601}{(2008)}}.

\bibitem{Fu2007b}
\textsc{L. Fu, C.~L. Kane} and \textsc{E.~J. Mele}, {\em {Topological
  Insulators in Three Dimensions}},
  {\href{http://dx.doi.org/10.1103/PhysRevLett.98.106803}{Phys. Rev. Lett.}}
  {\bf {\href{http://dx.doi.org/10.1103/PhysRevLett.98.106803}{98}}},
  {\href{http://dx.doi.org/10.1103/PhysRevLett.98.106803}{106803}}
  {\href{http://dx.doi.org/10.1103/PhysRevLett.98.106803}{(2007)}}.

\bibitem{Fu2007}
\textsc{L. Fu} and \textsc{C.~L. Kane}, {\em {Topological insulators with
  inversion symmetry}},
  {\href{http://dx.doi.org/10.1103/PhysRevB.76.045302}{Phys. Rev. B}} {\bf
  {\href{http://dx.doi.org/10.1103/PhysRevB.76.045302}{76}}},
  {\href{http://dx.doi.org/10.1103/PhysRevB.76.045302}{45302}}
  {\href{http://dx.doi.org/10.1103/PhysRevB.76.045302}{(2007)}}.

\bibitem{Qu2010}
\textsc{D.-X. Qu, Y.~S. Hor, J. Xiong, R.~J. Cava} and \textsc{N.~P. Ong}, {\em
  {Quantum Oscillations and Hall Anomaly of Surface States in the Topological
  Insulator Bi₂Te₃}},
  {\href{http://dx.doi.org/10.1126/science.1189792}{Science}} {\bf
  {\href{http://dx.doi.org/10.1126/science.1189792}{329}}},
  {\href{http://dx.doi.org/10.1126/science.1189792}{821}}
  {\href{http://dx.doi.org/10.1126/science.1189792}{(2010)}}.

\bibitem{Xiu2011}
\textsc{F. Xiu} {\it et~al.}\textsc{}, {\em {Manipulating surface states in
  topological insulator nanoribbons}},
  {\href{http://dx.doi.org/10.1038/nnano.2011.19}{Nat. Nano}} {\bf
  {\href{http://dx.doi.org/10.1038/nnano.2011.19}{6}}},
  {\href{http://dx.doi.org/10.1038/nnano.2011.19}{216}}
  {\href{http://dx.doi.org/10.1038/nnano.2011.19}{(2011)}}.

\bibitem{Koenig2007}
\textsc{M. K\"{o}nig, S. Wiedmann, C. Br\"{u}ne, A. Roth, H. Buhmann, L.~W.
  Molenkamp, X.-L. Qi} and \textsc{S.-C. Zhang}, {\em {Quantum Spin Hall
  Insulator State in HgTe Quantum Wells}},
  {\href{http://dx.doi.org/10.1126/science.1148047}{Science}} {\bf
  {\href{http://dx.doi.org/10.1126/science.1148047}{318}}},
  {\href{http://dx.doi.org/10.1126/science.1148047}{766}}
  {\href{http://dx.doi.org/10.1126/science.1148047}{(2007)}}.

\bibitem{Roth2009}
\textsc{A. Roth, C. Br\"{u}ne, H. Buhmann, L.~W. Molenkamp, J. Maciejko, X.-L.
  Qi} and \textsc{S.-C. Zhang}, {\em {Nonlocal Transport in the Quantum Spin
  Hall State}}, {\href{http://dx.doi.org/10.1126/science.1174736}{Science}}
  {\bf {\href{http://dx.doi.org/10.1126/science.1174736}{325}}},
  {\href{http://dx.doi.org/10.1126/science.1174736}{294}}
  {\href{http://dx.doi.org/10.1126/science.1174736}{(2009)}}.

\bibitem{Knez2011}
\textsc{I. Knez, R.-R. Du} and \textsc{G. Sullivan}, {\em {Evidence for Helical
  Edge Modes in Inverted InAs/GaSb Quantum Wells}},
  {\href{http://dx.doi.org/10.1103/PhysRevLett.107.136603}{Phys. Rev. Lett.}}
  {\bf {\href{http://dx.doi.org/10.1103/PhysRevLett.107.136603}{107}}},
  {\href{http://dx.doi.org/10.1103/PhysRevLett.107.136603}{136603}}
  {\href{http://dx.doi.org/10.1103/PhysRevLett.107.136603}{(2011)}}.

\bibitem{Zhou2008}
\textsc{B. Zhou, H.-Z. Lu, R.-L. Chu, S.-Q. Shen} and \textsc{Q. Niu}, {\em
  {Finite Size Effects on Helical Edge States in a Quantum Spin-Hall System}},
  {\href{http://dx.doi.org/10.1103/PhysRevLett.101.246807}{Phys. Rev. Lett.}}
  {\bf {\href{http://dx.doi.org/10.1103/PhysRevLett.101.246807}{101}}},
  {\href{http://dx.doi.org/10.1103/PhysRevLett.101.246807}{246807}}
  {\href{http://dx.doi.org/10.1103/PhysRevLett.101.246807}{(2008)}}.

\bibitem{Zhang2011}
\textsc{L.~B. Zhang, F. Cheng, F. Zhai} and \textsc{K. Chang}, {\em {Electrical
  switching of the edge channel transport in HgTe quantum wells with an
  inverted band structure}},
  {\href{http://dx.doi.org/10.1103/PhysRevB.83.081402}{Phys. Rev. B}} {\bf
  {\href{http://dx.doi.org/10.1103/PhysRevB.83.081402}{83}}},
  {\href{http://dx.doi.org/10.1103/PhysRevB.83.081402}{81402}}
  {\href{http://dx.doi.org/10.1103/PhysRevB.83.081402}{(2011)}}.

\bibitem{Krueckl2011b}
\textsc{V. Krueckl} and \textsc{K. Richter}, {\em {Switching Spin and Charge
  between Edge States in Topological Insulator Constrictions}},
  {\href{http://dx.doi.org/10.1103/PhysRevLett.107.086803}{Phys. Rev. Lett.}}
  {\bf {\href{http://dx.doi.org/10.1103/PhysRevLett.107.086803}{107}}},
  {\href{http://dx.doi.org/10.1103/PhysRevLett.107.086803}{86803}}
  {\href{http://dx.doi.org/10.1103/PhysRevLett.107.086803}{(2011)}}.

\bibitem{Dolcini2011}
\textsc{F. Dolcini}, {\em {Full electrical control of charge and spin
  conductance through interferometry of edge states in topological
  insulators}}, {\href{http://dx.doi.org/10.1103/PhysRevB.83.165304}{Phys. Rev.
  B}} {\bf {\href{http://dx.doi.org/10.1103/PhysRevB.83.165304}{83}}},
  {\href{http://dx.doi.org/10.1103/PhysRevB.83.165304}{165304}}
  {\href{http://dx.doi.org/10.1103/PhysRevB.83.165304}{(2011)}}.

\bibitem{Budich2012}
\textsc{J.~C. Budich, F. Dolcini, P. Recher} and \textsc{B. Trauzettel}, {\em
  {Phonon-Induced Backscattering in Helical Edge States}},
  {\href{http://dx.doi.org/10.1103/PhysRevLett.108.086602}{Phys. Rev. Lett.}}
  {\bf {\href{http://dx.doi.org/10.1103/PhysRevLett.108.086602}{108}}},
  {\href{http://dx.doi.org/10.1103/PhysRevLett.108.086602}{086602}}
  {\href{http://dx.doi.org/10.1103/PhysRevLett.108.086602}{(2012)}}.

\bibitem{Altshuler1980}
\textsc{B. Altshuler, D. Khmel'nitzkii, A. Larkin} and \textsc{P. Lee}, {\em
  {Magnetoresistance and Hall effect in a disordered two-dimensional electron
  gas}}, {\href{http://dx.doi.org/10.1103/PhysRevB.22.5142}{Phys. Rev. B}} {\bf
  {\href{http://dx.doi.org/10.1103/PhysRevB.22.5142}{22}}},
  {\href{http://dx.doi.org/10.1103/PhysRevB.22.5142}{5142}}
  {\href{http://dx.doi.org/10.1103/PhysRevB.22.5142}{(1980)}}.

\bibitem{Hikami1980}
\textsc{S. Hikami, A.~I. Larkin} and \textsc{Y. Nagaoka}, {\em {Spin--Orbit
  Interaction and Magnetoresistance in the Two-Dimensional Random System}},
  Prog. Theor. Phys. {\bf 63},  707  (1980).

\bibitem{Tkachov2011}
\textsc{G. Tkachov} and \textsc{E.~M. Hankiewicz}, {\em {Weak antilocalization
  in HgTe quantum wells and topological surface states: Massive versus massless
  Dirac fermions}}, {\href{http://dx.doi.org/10.1103/PhysRevB.84.035444}{Phys.
  Rev. B}} {\bf {\href{http://dx.doi.org/10.1103/PhysRevB.84.035444}{84}}},
  {\href{http://dx.doi.org/10.1103/PhysRevB.84.035444}{035444}}
  {\href{http://dx.doi.org/10.1103/PhysRevB.84.035444}{(2011)}}.

\bibitem{Lu2011PRL}
\textsc{H.-Z. Lu, J. Shi} and \textsc{S.-Q. Shen}, {\em {Competition between
  Weak Localization and Antilocalization in Topological Surface States}},
  {\href{http://dx.doi.org/10.1103/PhysRevLett.107.076801}{Phys. Rev. Lett.}}
  {\bf {\href{http://dx.doi.org/10.1103/PhysRevLett.107.076801}{107}}},
  {\href{http://dx.doi.org/10.1103/PhysRevLett.107.076801}{076801}}
  {\href{http://dx.doi.org/10.1103/PhysRevLett.107.076801}{(2011)}}.

\bibitem{Lu2011PRB}
\textsc{H.-Z. Lu} and \textsc{S.-Q. Shen}, {\em {Weak localization of bulk
  channels in topological insulator thin films}},
  {\href{http://dx.doi.org/10.1103/PhysRevB.84.125138}{Phys. Rev. B}} {\bf
  {\href{http://dx.doi.org/10.1103/PhysRevB.84.125138}{84}}},
  {\href{http://dx.doi.org/10.1103/PhysRevB.84.125138}{125138}}
  {\href{http://dx.doi.org/10.1103/PhysRevB.84.125138}{(2011)}}.

\bibitem{Garate2012}
\textsc{I. Garate} and \textsc{L. Glazman}, {\em {Weak Localization and
  Antilocalization in Topological Insulator Thin Films with Coherent
  Bulk-Surface Coupling}}, arXiv  1206.1239v1  (2012).

\bibitem{He2011}
\textsc{H.-T. He, G. Wang, T. Zhang, I.-K. Sou, G. Wong, J.-N. Wang, H.-Z. Lu,
  S.-Q. Shen} and \textsc{F.-C. Zhang}, {\em {Impurity Effect on Weak
  Antilocalization in the Topological Insulator Bi\_\{2\}Te\_\{3\}}},
  {\href{http://dx.doi.org/10.1103/PhysRevLett.106.166805}{Phys. Rev. Lett.}}
  {\bf {\href{http://dx.doi.org/10.1103/PhysRevLett.106.166805}{106}}},
  {\href{http://dx.doi.org/10.1103/PhysRevLett.106.166805}{166805}}
  {\href{http://dx.doi.org/10.1103/PhysRevLett.106.166805}{(2011)}}.

\bibitem{Liu2012b}
\textsc{M. Liu} {\it et~al.}\textsc{}, {\em {Crossover between Weak
  Antilocalization and Weak Localization in a Magnetically Doped Topological
  Insulator}}, {\href{http://dx.doi.org/10.1103/PhysRevLett.108.036805}{Phys.
  Rev. Lett.}} {\bf
  {\href{http://dx.doi.org/10.1103/PhysRevLett.108.036805}{108}}},
  {\href{http://dx.doi.org/10.1103/PhysRevLett.108.036805}{036805}}
  {\href{http://dx.doi.org/10.1103/PhysRevLett.108.036805}{(2012)}}.


\bibitem{Olshanetsky2010}
\textsc{E.~B. Olshanetsky, Z.~D. Kvon, G.~M. Gusev, N.~N. Mikhailov, S.~A.
  Dvoretsky} and \textsc{J.~C. Portal}, {\em {Weak antilocalization in HgTe
  quantum wells near a topological transition}},
  {\href{http://dx.doi.org/10.1134/S0021364010070052}{JETP Letters}} {\bf
  {\href{http://dx.doi.org/10.1134/S0021364010070052}{91}}},
  {\href{http://dx.doi.org/10.1134/S0021364010070052}{347}}
  {\href{http://dx.doi.org/10.1134/S0021364010070052}{(2010)}}.

\bibitem{Minkov2012}
\textsc{G.~M. Minkov, A.~V. Germanenko, O.~E. Rut, A.~A. Sherstobitov, S.~A.
  Dvoretski} and \textsc{N.~N. Mikhailov}, {\em {Weak antilocalization in HgTe
  quantum well with inverted energy spectrum}}, arXiv:1202.1093  (2012).

\bibitem{Iordanskii1994}
\textsc{S.~V. Iordanskii, Y.~B. Lyanda-Geller} and \textsc{G.~E. Pikus}, {\em
  {Weak localization in quantum wells with spin-orbit interaction}}, JETP Lett.
  {\bf 60},  206  (1994).

\bibitem{Knap1996}
\textsc{W. Knap} {\it et~al.}\textsc{}, {\em {Weak antilocalization and spin
  precession in quantum wells.}}, Phys. Rev. B {\bf 53},  3912  (1996).

\bibitem{Rothe2010}
\textsc{D.~G. Rothe, R.~W. Reinthaler, C.-X. Liu, L.~W. Molenkamp, S.-C. Zhang}
  and \textsc{E.~M. Hankiewicz}, {\em {Fingerprint of different spin-orbit
  terms for spin transport in HgTe quantum wells}},
  {\href{http://dx.doi.org/10.1088/1367-2630/12/6/065012}{New J. Phys.}} {\bf
  {\href{http://dx.doi.org/10.1088/1367-2630/12/6/065012}{12}}},
  {\href{http://dx.doi.org/10.1088/1367-2630/12/6/065012}{65012}}
  {\href{http://dx.doi.org/10.1088/1367-2630/12/6/065012}{(2010)}}.

\bibitem{Koenig2008}
\textsc{M. K\"{o}nig, H. Buhmann, L.~W. Molenkamp, T.~L. Hughes, C.-X. Liu,
  X.-L. Qi} and \textsc{S.-C. Zhang}, {\em {The Quantum Spin Hall Effect:
  Theory and Experiment}}, {\href{http://dx.doi.org/10.1143/JPSJ.77.031007}{J.
  Phys. Soc. Jpn.}} {\bf
  {\href{http://dx.doi.org/10.1143/JPSJ.77.031007}{77}}},
  {\href{http://dx.doi.org/10.1143/JPSJ.77.031007}{31007}}
  {\href{http://dx.doi.org/10.1143/JPSJ.77.031007}{(2008)}}.

\bibitem{Landauer1970}
\textsc{R. Landauer}, {\em {Electrical resistance of disordered one-dimensional
  lattices}}, {\href{http://dx.doi.org/10.1080/14786437008238472}{Philosophical
  Magazine}} {\bf {\href{http://dx.doi.org/10.1080/14786437008238472}{21}}},
  {\href{http://dx.doi.org/10.1080/14786437008238472}{863}}
  {\href{http://dx.doi.org/10.1080/14786437008238472}{(1970)}}.

\bibitem{Fisher1981}
\textsc{D.~S. Fisher} and \textsc{P.~A. Lee}, {\em {Relation between
  conductivity and transmission matrix}}, Phys. Rev. B {\bf 23},  6851  (1981).

\bibitem{Berry1984}
\textsc{M.~V. Berry}, {\em {Quantal phase factors accompanying adiabatic
  changes}}, {\href{http://dx.doi.org/10.1098/rspa.1984.0023}{Proc. R. Soc.
  Lond. A}} {\bf {\href{http://dx.doi.org/10.1098/rspa.1984.0023}{392}}},
  {\href{http://dx.doi.org/10.1098/rspa.1984.0023}{45}}
  {\href{http://dx.doi.org/10.1098/rspa.1984.0023}{(1984)}}.

\bibitem{Chang1996}
\textsc{M.-C. Chang} and \textsc{Q. Niu}, {\em {Berry phase, hyperorbits, and
  the Hofstadter spectrum: Semiclassical dynamics in magnetic Bloch bands.}},
  {\href{http://dx.doi.org/10.1103/PhysRevB.53.7010}{Phys. Rev. B}} {\bf
  {\href{http://dx.doi.org/10.1103/PhysRevB.53.7010}{53}}},
  {\href{http://dx.doi.org/10.1103/PhysRevB.53.7010}{7010}}
  {\href{http://dx.doi.org/10.1103/PhysRevB.53.7010}{(1996)}}.

\bibitem{krueckl2011a}
\textsc{V. Krueckl, M. Wimmer, I. Adagideli, J. Kuipers} and \textsc{K.
  Richter}, {\em {Weak Localization in Mesoscopic Hole Transport: Berry Phases
  and Classical Correlations}},
  {\href{http://dx.doi.org/10.1103/PhysRevLett.106.146801}{Phys. Rev. Lett.}}
  {\bf {\href{http://dx.doi.org/10.1103/PhysRevLett.106.146801}{106}}},
  {\href{http://dx.doi.org/10.1103/PhysRevLett.106.146801}{146801}}
  {\href{http://dx.doi.org/10.1103/PhysRevLett.106.146801}{(2011)}}.

\bibitem{Suzuura2002}
\textsc{H. Suzuura} and \textsc{T. Ando}, {\em {Crossover from Symplectic to
  Orthogonal Class in a Two-Dimensional Honeycomb Lattice}},
  {\href{http://dx.doi.org/10.1103/PhysRevLett.89.266603}{Phys. Rev. Lett.}}
  {\bf {\href{http://dx.doi.org/10.1103/PhysRevLett.89.266603}{89}}},
  {\href{http://dx.doi.org/10.1103/PhysRevLett.89.266603}{266603}}
  {\href{http://dx.doi.org/10.1103/PhysRevLett.89.266603}{(2002)}}.

\bibitem{McCann2006a}
\textsc{E. McCann, K. Kechedzhi, V.~I. Fal’ko, H. Suzuura, T. Ando} and
  \textsc{B.~L. Altshuler}, {\em {Weak-Localization Magnetoresistance and
  Valley Symmetry in Graphene}},
  {\href{http://dx.doi.org/10.1103/PhysRevLett.97.146805}{Phys. Rev. Lett.}}
  {\bf {\href{http://dx.doi.org/10.1103/PhysRevLett.97.146805}{97}}},
  {\href{http://dx.doi.org/10.1103/PhysRevLett.97.146805}{146805}}
  {\href{http://dx.doi.org/10.1103/PhysRevLett.97.146805}{(2006)}}.

\bibitem{Berry1987}
\textsc{M.~V. Berry} and \textsc{R.~J. Mondragon}, {\em {Neutrino billiards:
  time-reversal symmetry-breaking without magnetic fields}},
  {\href{http://dx.doi.org/10.1098/rspa.1987.0080}{Proc. R. Soc. Lond. A}} {\bf
  {\href{http://dx.doi.org/10.1098/rspa.1987.0080}{412}}},
  {\href{http://dx.doi.org/10.1098/rspa.1987.0080}{53}}
  {\href{http://dx.doi.org/10.1098/rspa.1987.0080}{(1987)}}.

\bibitem{Beenakker1997}
\textsc{C.~W.~J. Beenakker}, {\em {Random-matrix theory of quantum transport}},
  Rev. Mod. Phys. {\bf 69},  731  (1997).

\end{thebibliography}

\end{document}